\documentclass[conference]{IEEEtran}
\IEEEoverridecommandlockouts
\usepackage{cite}
\usepackage{amsmath,amssymb,amsfonts}
\usepackage{algorithmic}
\usepackage{graphicx}
\usepackage{textcomp}
\usepackage{xcolor}
\usepackage{amsfonts,amssymb}
\usepackage{multirow}
\usepackage{makecell}
\usepackage{array} 
\usepackage{longtable}
\usepackage{booktabs}
\usepackage{float}
\usepackage{diagbox}
\usepackage{subfigure}
\def\BibTeX{{\rm B\kern-.05em{\sc i\kern-.025em b}\kern-.08em
    T\kern-.1667em\lower.7ex\hbox{E}\kern-.125emX}}
\begin{document}

\title{Cooperative Task-Oriented Communication for Multi-Modal Data with Transmission Control}

\author{
	\IEEEauthorblockN{
		Siqi Wan\IEEEauthorrefmark{1}, 
		Qianqian Yang\IEEEauthorrefmark{2}, 
		Zhiguo Shi\IEEEauthorrefmark{2}, 
		Zhaohui Yang\IEEEauthorrefmark{2} 
		and Zhaoyang Zhang\IEEEauthorrefmark{2}} 
	\IEEEauthorblockA{\IEEEauthorrefmark{1}Polytechnic Institute, Zhejiang University, Hangzhou, China Email: 22260112@zju.edu.cn}
	\IEEEauthorblockA{\IEEEauthorrefmark{2}College of Information Science and Electronic Engineering, Zhejiang University, Hangzhou, China}\
    \IEEEauthorblockA{Emails: 22260112@zju.edu.cn, qianqianyang20@zju.edu.cn, shizg@zju.edu.cn,\\yang.zhaohui@kcl.ac.uk, ning\_ming@zju.edu.cn}
}

\maketitle

\begin{abstract}
Real-time intelligence applications in Internet of Things (IoT) environment depend on timely data communication. However, it is challenging to transmit and analyse massive data of various modalities. Recently proposed task-oriented communication methods based on deep learning have showed its superiority in communication efficiency. In this paper, we propose a cooperative task-oriented communication method for the transmission of multi-modal data from multiple end devices to a central server.  In particular, we use the transmission result of data of one modality, which is with lower rate, to control the transmission of other modalities with higher rate in order to reduce the amount of transmitted date. We take the human activity recognition (HAR) task in a smart home environment and design the semantic-oriented transceivers for the transmission of monitoring videos of different rooms and acceleration data of the monitored human. The numerical results demonstrate that by using the transmission control based on the obtained results of the received acceleration data, the transmission is reduced to 2\% of that without transmission control while preserving the performance on the HAR task.
\end{abstract}

\section{Introduction}
With the development of communication technology and the improvement in computing power of hardware, this is the era of Internet of everything. A large number of devices are connected to the communication network and massive heterogeneous data are generated anywhere and anytime. Intelligence Internet of Things (IoT) applications demand for low-latency and high-reliable data transmission. The traditional communication approach focusing on bit transmission seems insufficient to handle  this continuously growing demand. Semantic communication approaches, empowered by the advances in machine learning (ML) techniques, especially deep learning (DL), have achieved significant improvement in transmission efficiency than the transitional methods  is getting more attention\cite{weaver1953recent, bao2011towards}. 

Traditional communication system tragets for the accurate transmission of bit streams, while ignoring the semantic content and the effectiveness for the underlying tasks\cite{shannon1948mathematical}. However, semantic or task oriented communication systems end-to-end optimize semantic encoder and decoder based on DL to sent only the task-related semantic features over the channel. Effective semantic communication systems have been proposed for data transmission of text, image and speech in \cite{farsad2018deep, weng2021semantic, bourtsoulatze2019deep, han2022semantic, zhang2022wireless}. Other than these reconstruction tasks, DL-based semantic communication systems have been developed for different AI tasks, e.g. scene classification\cite{kang2022task} and object identification\cite{zhang2022multi}.

Most of the existing works on semantic communication consider the transmission of one modality between one pair of transmitter and receiver. However, a lot of IoT applications involve the transmission of multi-modal data, such as audio, video, sensing data, etc. from distributed IOT devices to a central server. Task-oriented semantic communication systems based on multi-modal data have been proposed in \cite{xie2022task, xie2021task}, which improves both the transmission efficiency and performance on the target tasks compared to traditional method.
Traditional communication strategies, e.g. hybrid automatic repeat request, have also applied in task-oriented communication systems to further improve communication reliability \cite{jiang2022deep, jiang2022wireless}. Inspired by these existing works, we proposed a cooperative task-oriented communication approach that exploits the semantic relationship between  different modalities to minimize the transmission cost without degrading the task performance in this paper. The contributions are summarized as follows:

\begin{figure*}[htb]
\centering
    \includegraphics[width=0.95\textwidth]{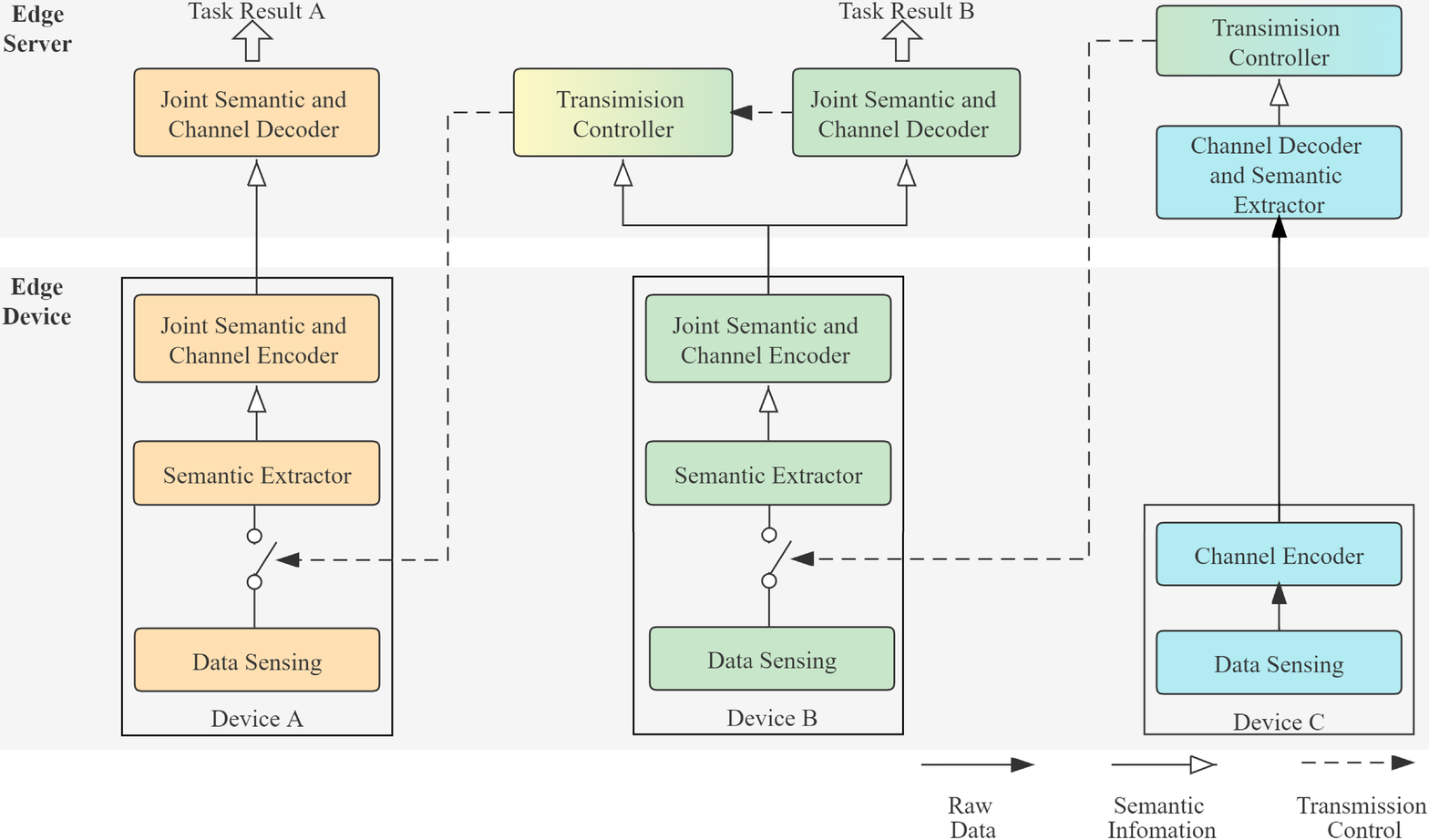}
\caption{System model of the task-oriented IoT communication framework based on edge-cloud architecture and multi-modal data fusion.}
\label{fig1}
\end{figure*}

\begin{itemize}
    \item We propose a task-oriented communication approach for the data transmission between distributed end devices that are sending data of different modalities and a central server. We introduction a transmission control mechanism where the transmission results of one modality are used to control the data transmission of another modality based on their semantic correlation.
    \item We take the human activity recognition (HAR) task for example and consider a smart home environment which consists of three cameras deployed in three different rooms, an accelerometer attached to the monitored human and a central server. We design a 3D-CNN based semantic transceiver for the transmission of monitoring videos from three cameras, and a random forest based semantic transceiver for the transmission of the acceleration date from the accelerometer. The posture results  obtained from acceleration data is used to control the uploading of the video semantic features.
    \item The simulation result demonstrates that the proposed communication framework achieves much lower communication overhead by exploiting semantic-oriented transmission and data transmission control mechanism without reducing the accuracy on the HAR task.
\end{itemize}

The rest of this paper is organized as follows. Section II introduces the system model of the proposed communication framework and gives an example of HAR task-oriented semantic communication framework in smart home. Section III  details the semantic communication model and transmission control mechanism for the HAR task. The simulation results are presented in section IV. Section V concludes this paper.

\section{System Model}

We consider a task-oriented communication system as shown in Fig.~\ref{fig1}, for multiple modal data transmission, where there are multiple devices in the system, each of which transmits data of different modalities to accomplish a common task jointly. The system consists of multiple transmitters, which are edge devices collecting and transmitting different types of data, and one receiver, which is a central server or the cloud server that performs the target task with the received data and sends out control signals to provoke transmission from the edge devices.  

1) At transmitter side, each transmitter collect data of different modalities for semantic extraction and encoding, which typically consists of three modules as shown in Fig.~\ref{fig1}. The \textbf{data sensing} module collects sensing signals to generate raw data. The \textbf{semantic extractor} module extracts semantic information from raw data and removes the irrelevant information exploiting semantic analysis algorithms such as deep learning method. The \textbf{joint semantic and channel encoder} module further compresses the extracted semantic information and generates a sequence of symbols to be transmitted over the physical channel. The semantic extractor and joint semantic and channel encoder modules are more computational heavy, and hence,  are usually installed on devices with sufficient computing source, such as intelligence cameras and smart speakers. Note that some edge devices have very low computational capacities, such as some simple sensors like infrared distance senors and thermometers. These devices are not capable of complex semantic extraction and encoding algorithm, and usually have low data rate, like temperature information. Hence, these devices only have the data sensing and channel encoder module, which directly encodes the sensed raw data and transmits to the receiver.

2) The receiver performs channel and semantic decoding of the received signals from each edge devices with the corresponding \textbf{joint channel and semantic decoder} module, which outputs the results of the given tasks. Additionally, the receiver exploits the \textbf{transmission controller} module, which uses the semantic information of one modality to control the data transmission of another modality based on their semantic relationship, in order to minimize the amount of transmitted data. Note that, some edge devices directly send the raw data to the server without any semantic extraction or processing due to the limited computational capability. For those received messages, the receiver also implements a \textbf{channel encoder and semantic extractor} module to reconstruct the raw data and extract the semantic information from it. Then a transmission controller is applied to decide the transmission of data from others edge devices based on the derived semantic information.

\begin{figure*}[htb]
\begin{center}
    \includegraphics[width=0.9\textwidth]{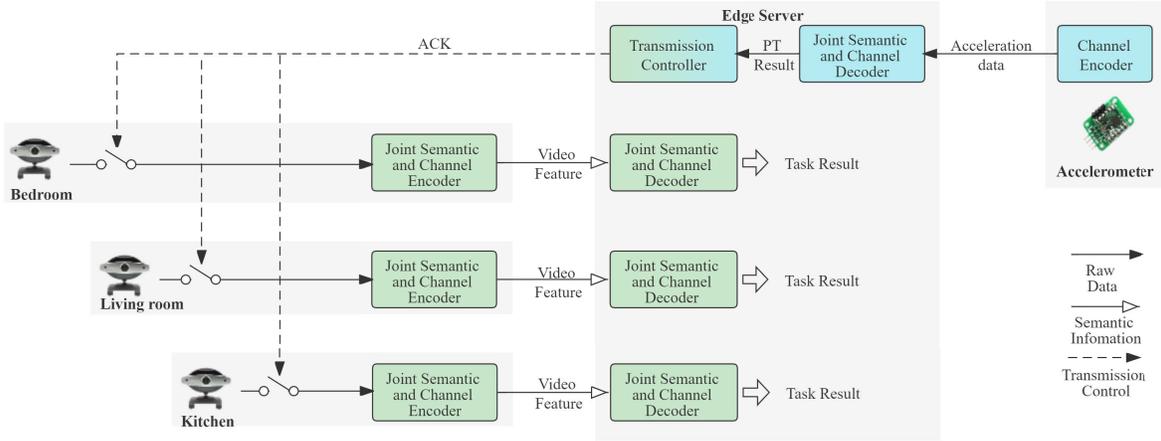}
\end{center}
\caption{Structure of the task-oriented semantic communication network.}
\label{fig2}
\end{figure*}

We adopt the proposed task-oriented communication framework for the HAR task in smart home, as shown in Fig.~\ref{fig2}. There are three cameras with certain level of computing sources installed in three different rooms, i.e., the bedroom, the living room and the kitchen, that record surveillance videos to monitor the human activities, and an accelerometer attached to the targeted person's left armpit to obtain the acceleration data of the monitored person. As explained above, each camera is implemented with a \textit{joint semantic and channel encoder} module that extracts and encodes human activities related semantic features from the recorded surveillance videos into transmitted symbols. Meanwhile, the accelerometer is a relatively small device with limited computing resource,  which is only able to collect and upload the raw data. The \textit{channel encoder} module at the accelerometer encodes the collected acceleration data into transmit symbols to be sent over the wireless channel. The received symbols at the cloud server are then decoded and analysed to obtain human postural transition (PT) results by the \textit{joint semantic and channel decoder} module. We note that when the human activity changes, the human posture must have changed. Taking account of this semantic relationship between human posture and activity, we utilize a \textit{transmission controller} at the server, which outputs a ACK signal to control the uploading of human activity semantic features from the three cameras. Only when a camera receives a positive ACK signal, it starts uploading the monitoring videos from this moment with the local \textit{joint semantic and channel encoder}. Then, the corresponding \textit{joint semantic and channel decoder} module at the server decodes and analyses the received features to generate the activity classification results. The details of the proposed system will be given in the next section.

\section{HAR-Oriented Communication Framework}
In this section, we present the details of the proposed HAR task-oriented system with transmission control, including the transceiver design for monitoring videos from three cameras to the server and the transceiver design for acceleration data from the accelerometer to the server.

\begin{figure*}[htb]
\begin{center}
    \includegraphics[width=0.9\textwidth]{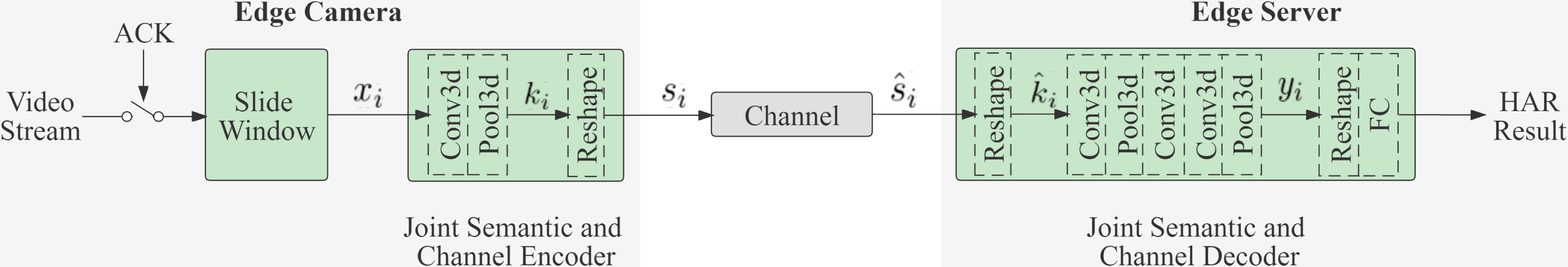}
\end{center}
\caption{Transceiver design for communication between the cameras and the server.}
\label{fig3}
\end{figure*}

\begin{figure*}[htb]
\begin{center}
    \includegraphics[width=0.9\textwidth]{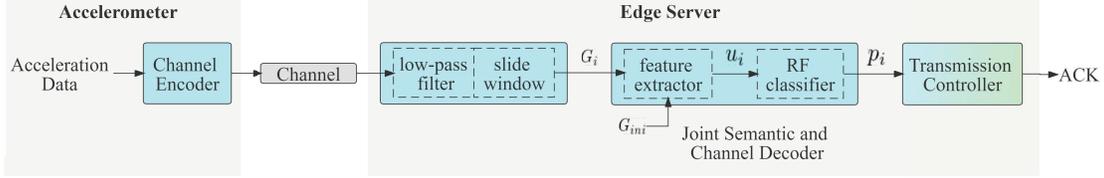}
\end{center}
\caption{Transceiver design for communication between the accelerometer and the server.}
\label{fig4}
\end{figure*}

\subsection{Semantic Transceiver for Monitoring Videos}
The semantic transceiver for monitoring videos transmission is illustrated in Fig.~\ref{fig3}, which includes a joint semantic and channel encoder at each camera and the corresponding joint semantic and channel decoder at the receiver. Each camera records monitoring videos at each room, which consists of continuous image frames. It then decide whether to further process recorded video by the recevied ACK signal sent out by the cloud server. If the camera has received a positive ACK signal, it sends a video segment from this moment consisting of 16 consecutive image frames, sampled by a slide window, which is denoted by a four dimensional matrix, $x_i \in \mathbb{R}^{16\times112\times112\times3}$, where $i$ denote the index of the current sample and the last dimension represents three channel of RGB images. 
Then, this matrix $x_i$ is fed into the joint semantic and channel encoder, which consists of a three dimensional (3D) convolutional layer with a kernel of size $(3,3,3)$ and $(1,1,1)$ padding, a max-pooling layer with a $(5,5,3)$ kernel with $(5,5,3)$ striding, which generates a video semantic feature matrix denoted as $k_i \in \mathbb{R}^{4\times5\times22\times22}$\cite{tran2015learning}. 
Finally, a reshape layer as the channel encoder maps the high dimensional semantic features into a semantic symbol sequence denoted as $s_i \in \mathbb{R}^{4840\times2}$ where the first and second channels correspond to the real and imaginary parts of the complex signals respectively. 

The symbols, $s_i$, is sent over a physical channel to the central server. The received semantic symbols are denoted by $\hat{s_i} = s_i + n$, 
where $\hat{s}_i \in \mathbb{R}^{4840\times2}$ and $n\sim N(0,\sigma^2)$ represents the addictive white Gaussian noise (AWGN) with zero mean and a variance of $\sigma$. The signal-noise ratio (SNR) in dB of the physical channel is given by
\begin{equation}
    SNR = 10 \log_{10} \big( \frac{P}{\sigma^2} \big),
    \label{eq_SNR}
\end{equation}
where $P$ is the power of the transmit symbols.

The received symbols at the central server, $\hat{s}_i$, is firstly resized into a 4 dimensional semantic feature matrix denoted by $\hat{k}_i \in \mathbb{R}^{4\times5\times22\times22}$ by a reshape layer, which corresponds to the process of channel decoding. Then, a 3D-CNN consisting of three 3D convolutional layers with $(3,3,3)$ kernel and $(1,1,1)$ padding, and two 3D max-pooling layers with $(2,2,2)$ kernel and $(2,2,2)$ striding\cite{tran2015learning}, is applied to obtain the deeper semantic feature, $y_i \in \mathbb{R}^{8\times1\times5\times5}$. 
We consider the HAR classification as the target task, the goal of which is to identify 5 types of recognizable daily activities, including:
\begin{itemize}
    \item Sleeping: user lies on the bed in the bedroom.
    \item Resting: user sits in the sofa in the living room, reading or watching TV.
    \item Dress-up: user dresses or undresses in front of the closet in bedroom.
    \item Eating: user sits at the table or eats in the kitchen.
    \item Calling: user sits at the desk and makes a phone call in the living room.
\end{itemize}
To obtain this HAR claasification result, the high dimensional semantic feature is resized by a reshape layer into a vector and then a fully connected linear layer to output the probabilities of all 5 types of activity. The activity with the highest probability is the final HAR classification result.

\subsection{Transceiver for Acceleration data}

Although exploiting the video data is able to obtain accurate prediction results on the HAR task with all required information, the transmission cost of continuous video data is high. Note that every human activity involves a series of PTs so a PT suggests the human activity may change. We utilize acceleration data collected by accelerometer worn on the user to detect PTs. When a PT is detected, the transmission controller at the server sends out positive ACK signal to the corresponding camera to activate the transmission of monitoring videos. We choose 4 types of human daily postures, i.e. lying, sitting, standing and walking to recognize, in order to support the classification of the above 5 types of daily activities. 

Fig.~\ref{fig4} illustrates the  the transceiver design for the transmission of acceleration data. The accelerometer collects 3-axis acceleration analog signals and an ADC converts these signals into digital signals which is directly encoded for wireless transmission by a channel encoder without any further preprocessing due to the limited computing capacity of the accelerometer. 

After recovered by the channel decoder at the receiver, the 3-axis acceleration signals are input to a low-pass filter of 0.3Hz cutoff frequency to obtain the low-frequency acceleration signals which provides the user’s movement orientation by approximating the gravitational acceleration \cite{curone2010real}. The continuous 3-axis low-frequency acceleration signals are then sampled by the 1s slide window to obtain a gravitational acceleration matrix, denoted by $G_i$, where $G_i\in\mathbb{R}^{3\times50}$,  where $i$ is the time sequence number. Let $g_{ij}=\in\mathbb{R}^{3}$ denote the $j$-th set of 3-axis values in $G_i$. We assume $g_{def}=[0,0,1]\in \mathbb{R}^3$ as the default gravity vector presenting the default movement orientation. 

Then, the feature extractor calculates the $i$-th cosine values, $u_i$, by the angle between $g_{def}$ and each vector $g_{ij}$ in $G_i$, which presents the movement orientation transition degree, given by 
\begin{equation}
    u_i = \frac{1}{50} \sum_{j=1}^{50} \frac{g_{ij}\cdot g_{def}}{|g_{ij}| \cdot |g_{def}|}.
\end{equation}
Different postures can be distinguished by the movement orientation transition degree. We use a random forest (RF) based classifier to generate posture classification result, $p_i$ with $u_i$ as the input. 

We then input $p_i$ into the transmission controller to detect PT by comparing the current posture classification result $p_i$ and the previous result $p_{i-1}$. To avoid false detection, a PT is valid only when the posture $p_i$ lasts for more than 2 seconds after the PT is detected. 
When the PT is valid, the transmission controller sends out a positive $ACK$ signals to the corresponding edge cameras as a command for uploading video semantic features.

\section{Simulation}
\subsection{Setup}
We adopt a daily activity dataset for simulation in which videos, audios and digital data including acceleration and temperature are collected simultaneously by sensors installed in a smart home environment or attached to the monitored human's body\cite{fleury2009svm}. 
We select two modalities of data from the dataset, i.e. video and 3-axis acceleration data, for the simulation. The videos are recorded by three cameras installed in bedroom, living room and kitchen respectively and the 3-axis acceleration data are collected by an accelerometer worn on user's left armpit.

For the collection of training data, we collect the video clips of daily activity from the raw video and the corresponding acceleration data. We apply the proposed video sampling and posture feature extraction methods on the video clips and acceleration raw data respectively to generate 14,199 video samples and 1,242 posture samples. For human posture classification task, labels of posture samples include ``lying'', ``sitting'', ``satnding'' and ``walking''. For HAR task, video samples are labeled as five types corresponding to human daily activities in smart home, i.e. ``sleeping'', ``resting'', ``dress-up'', ``eating'' and ``calling'', as aforementioned. All labels are encoded in one-hot format. We split the samples into training set and test set in a ratio of 7:3.

We implement the proposed method on Pytorch library. The semantic transceiver for monitoring video trained on the video dataset with the cross-entropy loss function. The learning rate is set as 0.003 initially and is divided by 4 every 4 epochs. The batchsize is 32 and the training is stopped after 15 epochs. The posture classification uses RF model and is optimized on the acceleration dataset.

\subsection{Performance}
\textit{1) Communication Overhead:} 
We compare the communication overhead among three video compression methods, i.e. the proposed task-oriented semantic communication framework (HAR-SC), the proposed HAR-SC combined with transmission control (HAR-SC-TC) and the MPEG-4. 

In HAR-SC, the communication overhead, $C$, is defined as the number of transmitted symbols required for the target task. The communication overhead of HAR-SC, $C_{SC}$, is calculated through
$$
C_{SC} = L \times N_f
$$
where $L$ is the length of the transmitted symbol vector and $N_f$ is the total number of symbol vectors required for the target task. 

In HAR-SC-TC, the video semantic features are only uploaded when a valid PT is detected and the communication overhead, $C_{TC}$, is calculated by
$$
C_{TC} = L \times N_t
$$
where $N_t$ is the number of feature uploading required for the target task.

Besides, we assume the channel bandwidth as 1Hz and denote the communication overhead of MPEG-4, $C_{MPEG-4}$, as
$$
C_{MPEG} = \frac{N_b}{\log_2{(1+10^{\gamma/10})}}
$$
where $N_b$ is the data size of video compressed by MEPG-4 encoder in bit and $\gamma$ is the SNR of the AWGN channel in dB.

We use a 110MB video consisting of 29,640 frames from which 1,852 semantic features are extracted through the proposed video sampling method. The calculation results are shown in Table.~\ref{tab1}. The communication overhead of HAR-SC is reduced by 97.5$\%$, 95.7$\%$, 93.8$\%$ and 91.9$\%$ compared with MPEG-4 scheme with channel SNR equal to 7dB, 13dB, 19dB and 25dB. The communication overhead of HAR-SC-TC is reduced by 98.0$\%$ compared with HAR-SC. It proves that the proposed task-oriented communication framework reduces the communication overhead significantly.
\begin{table}[tb]
    \renewcommand{\arraystretch}{1.5}
    \caption{The communication overhead under different compression methods}
    \label{tab1}
    \centering
    \begin{tabular}{c|c}
        \hline
        method         &communication overhead
        \\ \hline
        MPEG-4         &\makecell[c]{356,573,829(SNR=7dB)\\[0.2ex]
                                      210,237,977(SNR=13dB)\\[0.2ex]
                                      145,780,221(SNR=19dB)\\[0.2ex]
                                      111,048,888(SNR=25dB)}
        \\[0.2ex] \hline
        HAR-SC         & 8,963,680
        \\ \hline
        HAR-SC-TC      & 174,240
        \\ \hline
    \end{tabular}
\end{table}

\begin{figure}[tb]
\begin{center}
    \includegraphics[width=0.43\textwidth]{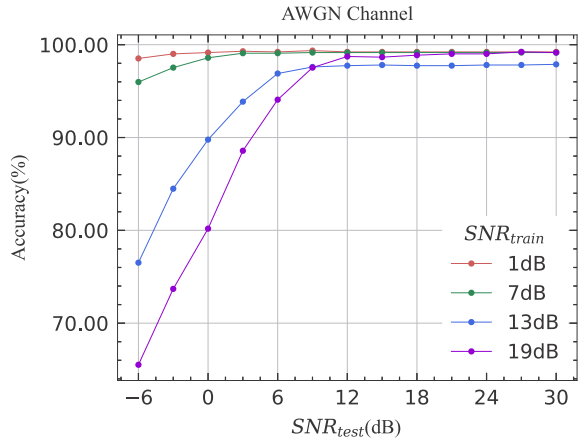}
\end{center}
\caption{Classification accuracy of HAR-SC over an AWGN channel.}
\label{fig5}
\end{figure}

\textit{2) Classification Accuracy:} 
We then evaluate the performance  of HAR-SC and HAR-SC-TC on the HAR task in terms of activity classification accuracy. For HAR-SC, the 3D-CNN based semantic transceiver is trained under different channel conditions where the SNRs are calculated via (\ref{eq_SNR}). The optimized semantic transceiver is tested under SNR equal to 7dB, 13dB, 19dB and 25dB respectively and the results are shown in Fig.~\ref{fig5}. 
Note that the accuracy drops when the $SNR_{test}$ is much lower than the $SNR_{train}$ because the transceiver is not able to to correct the transmission error when the channel noise is too big that it has not been trained for 
We also note that the $SNR_{test}$ is around or higher than $SNR_{train}$, the accuracy tends to reach a saturation value which depends on the model training. The simulation result also shows that the transceiver optimized under the worse channel condition has better performance under poor channel condition, and is more stable when channel condition changes.

For HAR-SC-TC, we test the 3D-CNN based semantic transceiver  combined with the RF based transceiver for acceleration data on the considered HAR task in terms of classification accuracy. The results are shown in Table.~\ref{tab3}, where the value is the number of times that an activity is detected in a room and the value in the following bracket is the classification accuracy. 
The frequency of uploading in HAR-SC-TC is much lower than HAR-SC, which means the communication overhead and power consumption are reduced remarkably. Besides, the classification accuracy of HAR-SC-TC is as high as HAR-SC, which proves the transmission controller module reduce the communication overhead without degrading the performance on target task. We remind that the semantic transceiver for monitoring video and the transceiver for acceleration data in HAR-SC-TC are designed and optimized separately. This results validate the effectiveness of using PT results predicted using the accelatation data as transmssion control signals to decide the tranmission of video signals.

\begin{table}[tb]
    \renewcommand{\arraystretch}{1.5}
    \caption{The classification results of HAR-SC-TC}
    \label{tab3}
    \centering
    \begin{tabular}{c|c|c|c}
        \hline
         \diagbox[]{activity}{room} & living room   & bedroom       & kitchen \\ \hline
         sleep                   & 0             & 10(100$\%$)   & 0       \\ \hline
         rest                    & 0             & 12(100$\%$)   & 0       \\ \hline
         dress-up                & 0             & 2(100$\%$)    & 0       \\ \hline
         eat                    & 0             & 0             & 1(100$\%$) \\ \hline
         call                   & 13(100$\%$)   & 0             & 0         \\ \hline
    \end{tabular}
\end{table}

\section{Conclusion}
In this paper, we proposed a task-oriented communication method for the transmission of multi-modal data from multiple end devices to a central server. In particular, we proposed to use the received information of data of one modality to control the transmission of data of another modality which is with much higher data rate. We took the HAR task in a smart home environment for example. More specifically, we designed a 3D-CNN based transceiver for the semantic communication of monitoring videos, which outputs the HAR results, and a RF-based transceiver for the transmission of acceleration data, which outputs the PT results. The PT results are used as the tranmission control signals that decide whether to transmit the monitoring videos to the server. Numerical results demonstrates that the proposed achieves a much lower communication cost than traditional video encoder without degrading the performance on the target task, which validates the effectiveness of the proposed tranmission control.

\bibliographystyle{IEEEtran}
\bibliography{IEEEabrv,myrefs}

\begin{thebibliography}{10}
\providecommand{\url}[1]{#1}
\csname url@samestyle\endcsname
\providecommand{\newblock}{\relax}
\providecommand{\bibinfo}[2]{#2}
\providecommand{\BIBentrySTDinterwordspacing}{\spaceskip=0pt\relax}
\providecommand{\BIBentryALTinterwordstretchfactor}{4}
\providecommand{\BIBentryALTinterwordspacing}{\spaceskip=\fontdimen2\font plus
\BIBentryALTinterwordstretchfactor\fontdimen3\font minus
  \fontdimen4\font\relax}
\providecommand{\BIBforeignlanguage}[2]{{%
\expandafter\ifx\csname l@#1\endcsname\relax
\typeout{** WARNING: IEEEtran.bst: No hyphenation pattern has been}%
\typeout{** loaded for the language `#1'. Using the pattern for}%
\typeout{** the default language instead.}%
\else
\language=\csname l@#1\endcsname
\fi
#2}}
\providecommand{\BIBdecl}{\relax}
\BIBdecl

\bibitem{weaver1953recent}
W.~Weaver, ``Recent contributions to the mathematical theory of
  communication,'' \emph{ETC: a review of general semantics}, pp. 261--281,
  1953.

\bibitem{bao2011towards}
J.~Bao, P.~Basu, M.~Dean, C.~Partridge, A.~Swami, W.~Leland, and J.~A. Hendler,
  ``Towards a theory of semantic communication,'' in \emph{2011 IEEE Network
  Science Workshop}, 2011, pp. 110--117.

\bibitem{shannon1948mathematical}
C.~E. Shannon, ``A mathematical theory of communication,'' \emph{The Bell
  system technical journal}, vol.~27, no.~3, pp. 379--423, 1948.

\bibitem{farsad2018deep}
N.~Farsad, M.~Rao, and A.~Goldsmith, ``Deep learning for joint source-channel
  coding of text,'' in \emph{2018 IEEE International Conference on Acoustics,
  Speech and Signal Processing (ICASSP)}, 2018, pp. 2326--2330.

\bibitem{weng2021semantic}
Z.~Weng and Z.~Qin, ``Semantic communication systems for speech transmission,''
  \emph{IEEE Journal on Selected Areas in Communications}, vol.~39, no.~8, pp.
  2434--2444, 2021.

\bibitem{bourtsoulatze2019deep}
E.~Bourtsoulatze, D.~Burth~Kurka, and D.~Gündüz, ``Deep joint source-channel
  coding for wireless image transmission,'' \emph{IEEE Transactions on
  Cognitive Communications and Networking}, vol.~5, no.~3, pp. 567--579, 2019.

\bibitem{han2022semantic}
T.~Han, Q.~Yang, Z.~Shi, S.~He, and Z.~Zhang, ``Semantic-preserved
  communication system for highly efficient speech transmission,'' \emph{IEEE
  Journal on Selected Areas in Communications}, vol.~41, no.~1, pp. 245--259,
  2023.

\bibitem{zhang2022wireless}
Z.~Zhang, Q.~Yang, S.~He, M.~Sun, and J.~Chen, ``Wireless transmission of
  images with the assistance of multi-level semantic information,'' in
  \emph{2022 International Symposium on Wireless Communication Systems
  (ISWCS)}, 2022, pp. 1--6.

\bibitem{kang2022task}
X.~Kang, B.~Song, J.~Guo, Z.~Qin, and F.~R. Yu, ``Task-oriented image
  transmission for scene classification in unmanned aerial systems,''
  \emph{IEEE Transactions on Communications}, vol.~70, no.~8, pp. 5181--5192,
  2022.

\bibitem{zhang2022multi}
Y.~Zhang, W.~Xu, H.~Gao, and F.~Wang, ``Multi-user semantic communications for
  cooperative object identification,'' in \emph{2022 IEEE International
  Conference on Communications Workshops (ICC Workshops)}, 2022, pp. 157--162.

\bibitem{xie2022task}
H.~Xie, Z.~Qin, X.~Tao, and K.~B. Letaief, ``Task-oriented multi-user semantic
  communications,'' \emph{IEEE Journal on Selected Areas in Communications},
  vol.~40, no.~9, pp. 2584--2597, 2022.

\bibitem{xie2021task}
H.~Xie, Z.~Qin, and G.~Y. Li, ``Task-oriented multi-user semantic
  communications for vqa,'' \emph{IEEE Wireless Communications Letters},
  vol.~11, no.~3, pp. 553--557, 2022.

\bibitem{jiang2022deep}
P.~Jiang, C.-K. Wen, S.~Jin, and G.~Y. Li, ``Deep source-channel coding for
  sentence semantic transmission with harq,'' \emph{IEEE Transactions on
  Communications}, 2022.

\bibitem{jiang2022wireless}
{Jiang, Peiwen and Wen, Chao-Kai and Jin, Shi and Li, Geoffrey Ye}, ``Wireless
  semantic communications for video conferencing,'' \emph{IEEE Journal on
  Selected Areas in Communications}, vol.~41, no.~1, pp. 230--244, 2023.

\bibitem{tran2015learning}
D.~Tran, L.~Bourdev, R.~Fergus, L.~Torresani, and M.~Paluri, ``Learning
  spatiotemporal features with 3d convolutional networks,'' in
  \emph{Proceedings of the IEEE international conference on computer vision},
  2015, pp. 4489--4497.

\bibitem{curone2010real}
D.~Curone, G.~M. Bertolotti, A.~Cristiani, E.~L. Secco, and G.~Magenes, ``A
  real-time and self-calibrating algorithm based on triaxial accelerometer
  signals for the detection of human posture and activity,'' \emph{IEEE
  Transactions on Information Technology in Biomedicine}, vol.~14, no.~4, pp.
  1098--1105, 2010.

\bibitem{fleury2009svm}
A.~Fleury, M.~Vacher, and N.~Noury, ``Svm-based multimodal classification of
  activities of daily living in health smart homes: Sensors, algorithms, and
  first experimental results,'' \emph{IEEE Transactions on Information
  Technology in Biomedicine}, vol.~14, no.~2, pp. 274--283, 2010.

\end{thebibliography}

\end{document}